\documentclass[12pt,a4paper]{article}
\usepackage[T1]{fontenc}
\usepackage[latin1]{inputenc}
\usepackage{graphicx}
\usepackage{setspace}
\usepackage[pagewise]{lineno}
\usepackage{amsmath}
\usepackage{cite}
\usepackage{geometry}
\usepackage{booktabs}
\usepackage[table,xcdraw]{xcolor}
\usepackage[normalem]{ulem}
\useunder{\uline}{\ul}{}

\usepackage[latin1]{inputenc}                                 
\usepackage{color}                                            
\usepackage{array}                                            
\usepackage{longtable}                                        
\usepackage{calc}                                             
\usepackage{multirow}                                         
\usepackage{hhline}                                           
\usepackage{ifthen}

\title{Please, no more scientific journals!\\The strategy of the scientific publication system\\
\vspace{1cm}}

\author{Miguel A. Fortuna\\miguel.fortuna@ieu.uzh.ch\\\\\\
Department of Evolutionary Biology and Environmental Studies\\University of Zurich, Zurich (Switzerland)}

\vspace{3cm}
\date{}

\begin{document}
\maketitle
\baselineskip=8.5 mm

\begin{center}
{\bf Abstract.}
\end{center}

\singlespacing
\noindent In the same way ecosystems tend to increase maturity by decreasing the flow of energy per unit biomass, we should move towards a more mature science by publishing less but high-quality papers and getting away from joining large teams in small roles. That is, we should decrease our scientific productivity for good.

\doublespacing
\newpage
The number of scientific journals listed in the Journal of Citation Reports has increased during the last decade from $6,443$ to $8,860$ (Table 1), and hence, the world's scientific production (i.e., number of papers published by the scientific community per year) has grown from $905,349$ to $1,398,003$ in that period (Fig. 1a). Journals act like fertilizers for the soil, opening up new opportunities for scientists to publish the results of their research. Consolidated scientists can increase their scientific productivity (i.e., number of papers published per scientist and year). On the other hand, novel scientists---who did not publish before because of the fierce competition for space in journals---can enter into the scientific publication system. Disentangling the detrimental consequences of the never-ending increase of the number of journals (i.e., higher scientific productivity) from its advantageous effects (i.e., recruiting novel scientists) is of paramount importance to support or discourage the proliferation of new scientific journals.

Since the number of authors has increased (from $3,052,068$ to $5,318,208$) faster than the number of papers published during the last decade ($75\%$ and $50\%$, respectively), adding more journals into the publication system seems to have reduced scientific productivity. Note, however, that using the ratio $\frac{\text{\#papers}}{\text{\#authors}}$ (i.e,. scientific production over the number of authors) as a proxy of scientific productivity (Fig. 1b) assumes single-authored papers. This negative linear relationship still holds for multiple-author papers if we assume that the average number of authors signing a research paper has not changed over time. Yet, the only way to keep scientific productivity constant over time when the number of authors increases faster than the number of papers is by increasing the number of authors signing a paper. This suggests that, even when the number of papers grows at a lower pace than the number of authors, the higher the frequency of papers signed by many authors is, the higher might be scientific productivity.

The relative frequency of papers signed by five or more authors has increased during the last decade from $40\%$ to $53\%$ (Fig. 1c). Single-author papers, papers signed by two, three and four authors---but not by five or more authors---have decreased over time (shown by negative and statistically significant correlations; see Table 2). This resulting increase in the average number of authors per paper over time must be incorporated in any measure of scientific productivity.

One way to measure scientific productivity per year taking into account the increase of the frequency of papers signed by many authors is the following:
\begin{equation*}
  \text{scientific productivity}=\frac{\sum\limits_{i=1}^{20}\Big(\text{\#papers}_{(i)} *\hspace{0.1cm}i\Big)}{\text{\#authors}},
\vspace{-0.05cm}
\end{equation*}

\noindent where $\text{\#papers}_{(i)}$ is the number of papers signed by $i$ authors and $\text{\#authors}$ is the total number of authors of all the papers published that year. Note that we did not consider papers signed by more than 20 authors because the percentage of those papers was lower than $0.5\%$. This measure shows that scientific productivity has increased over time ($5\%$ during the last decade; Fig. 1d) as a consequence of the proliferation of new journals.

While it is true that having more journals has allowed novel scientists to enter into the scientific publication system, most of them are just part of multi-authored papers. In the same way ecosystems tend to increase maturity by decreasing the flow of energy per unit biomass (Margalef, {\em The American Naturalist}, 97 (1963): 357-374), we need to move towards a more mature science where we should publish less but high-quality papers (i.e., trend towards decreasing scientific productivity). It's time to face the conflict between the strategies of science and publishers.

\newpage
\begin{figure}[h!]
  \begin{center}
    \centerline{\includegraphics[width=0.42\columnwidth]{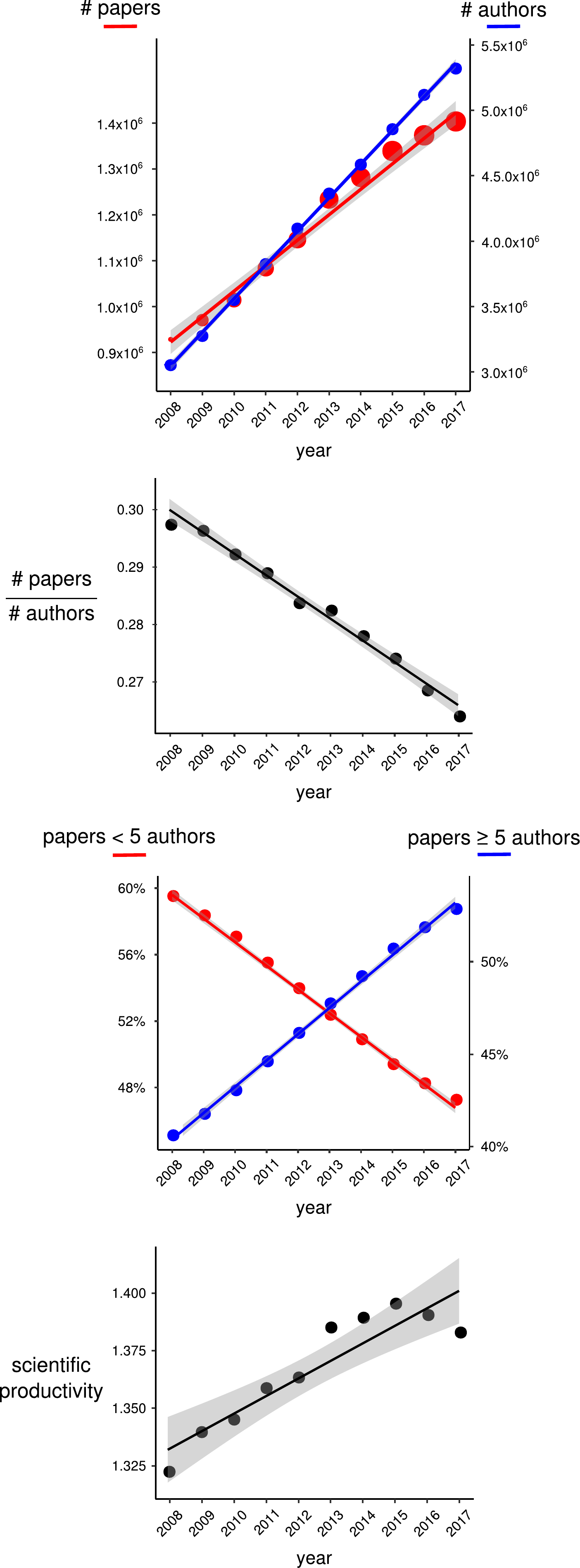}}
  \end{center}
\end{figure}
\vspace{-1.5 cm}
\singlespacing
\noindent Figure 1. {\bf The strategy of the scientific publication system.} a) Trends in the number of papers (red) and authors (blue) over the last decade. b) Scientific productivity over time measured as the ratio between the number of papers published per year and the number of scientists authored them. c) Change in the frequency of papers signed by four or less scientists (red) and more than four scientists (blue). {\bf d)} Scientific productivity over time measured after taking into account the increase in the number of authors per paper.

\newpage
\section*{Data.}
\singlespacing
Table 1. {\bf Summary descriptors of the scientific publication system in the last decade.} Number of journals included in the Science Citation Index Expanded, number of scientific articles, number of distinct authors signing those articles, articles signed by less than five authors, and articles signed by five or more authors.
\begin{center}
  {\renewcommand\normalsize{\scriptsize}
    \normalsize
\def\ifundefined#1{\expandafter\ifx\csname#1\endcsname\relax}

\ifundefined{inputGnumericTable}

	\def\gnumericTableEnd{\end{document}}


	\documentclass[12pt%
                    ]{report}
       \usepackage[latin1]{inputenc}
       \usepackage{fullpage}
       \usepackage{color}
       \usepackage{array}
       \usepackage{longtable}
       \usepackage{calc}
       \usepackage{multirow}
       \usepackage{hhline}
       \usepackage{ifthen}

	\begin{document}

\else

    \def\gnumericTableEnd{}


\fi


\providecommand{\gnumericmathit}[1]{#1} 
\providecommand{\gnumericPB}[1]%
{\let\gnumericTemp=\\#1\let\\=\gnumericTemp\hspace{0pt}}
 \ifundefined{gnumericTableWidthDefined}
        \newlength{\gnumericTableWidth}
        \newlength{\gnumericTableWidthComplete}
        \newlength{\gnumericMultiRowLength}
        \global\def\gnumericTableWidthDefined{}
 \fi
 \ifthenelse{\isundefined{\languageshorthands}}{}{\languageshorthands{english}}
\providecommand\gnumbox{\makebox[0pt]}

\setlength{\bigstrutjot}{\jot}
\setlength{\extrarowheight}{\doublerulesep}

\setlongtables

\setlength\gnumericTableWidth{%
	53pt+%
	85pt+%
	80pt+%
	92pt+%
	107pt+%
	96pt+%
	91pt+%
0pt}
\def\gumericNumCols{7}
\setlength\gnumericTableWidthComplete{\gnumericTableWidth+%
         \tabcolsep*\gumericNumCols*2+\arrayrulewidth*\gumericNumCols}
\ifthenelse{\lengthtest{\gnumericTableWidthComplete > \linewidth}}%
         {\def\gnumericScale{\ratio{\linewidth-%
                        \tabcolsep*\gumericNumCols*2-%
                        \arrayrulewidth*\gumericNumCols}%
{\gnumericTableWidth}}}%
{\def\gnumericScale{1}}


\ifthenelse{\isundefined{\gnumericColA}}{\newlength{\gnumericColA}}{}\settowidth{\gnumericColA}{\begin{tabular}{@{}p{53pt*\gnumericScale}@{}}x\end{tabular}}
\ifthenelse{\isundefined{\gnumericColB}}{\newlength{\gnumericColB}}{}\settowidth{\gnumericColB}{\begin{tabular}{@{}p{85pt*\gnumericScale}@{}}x\end{tabular}}
\ifthenelse{\isundefined{\gnumericColC}}{\newlength{\gnumericColC}}{}\settowidth{\gnumericColC}{\begin{tabular}{@{}p{80pt*\gnumericScale}@{}}x\end{tabular}}
\ifthenelse{\isundefined{\gnumericColD}}{\newlength{\gnumericColD}}{}\settowidth{\gnumericColD}{\begin{tabular}{@{}p{92pt*\gnumericScale}@{}}x\end{tabular}}
\ifthenelse{\isundefined{\gnumericColE}}{\newlength{\gnumericColE}}{}\settowidth{\gnumericColE}{\begin{tabular}{@{}p{107pt*\gnumericScale}@{}}x\end{tabular}}
\ifthenelse{\isundefined{\gnumericColF}}{\newlength{\gnumericColF}}{}\settowidth{\gnumericColF}{\begin{tabular}{@{}p{96pt*\gnumericScale}@{}}x\end{tabular}}
\ifthenelse{\isundefined{\gnumericColG}}{\newlength{\gnumericColG}}{}\settowidth{\gnumericColG}{\begin{tabular}{@{}p{91pt*\gnumericScale}@{}}x\end{tabular}}

\begin{longtable}[c]{%
	b{\gnumericColA}%
	b{\gnumericColB}%
	b{\gnumericColC}%
	b{\gnumericColD}%
	b{\gnumericColE}%
	b{\gnumericColF}%
	b{\gnumericColG}%
	}






	 \gnumericPB{\centering}\gnumbox{\textbf{year}}
	&\gnumericPB{\centering}\gnumbox{\textbf{\#journals}}
	&\gnumericPB{\centering}\gnumbox{\textbf{\#articles}}
	&\gnumericPB{\centering}\gnumbox{\textbf{\#authors}}
	&\gnumericPB{\centering}\gnumbox{\textbf{< 5 authors}}
	&\gnumericPB{\centering}\gnumbox{\textbf{$\ge$ 5 authors}}
\\
	 \gnumericPB{\centering}\gnumbox{2008}
	&\gnumericPB{\centering}\gnumbox{6443}
	&\gnumericPB{\centering}\gnumbox{905349}
	&\gnumericPB{\centering}\gnumbox{3052068}
	&\gnumericPB{\centering}\gnumbox{59.45\%}
	&\gnumericPB{\centering}\gnumbox{40.55\%}
\\
	 \gnumericPB{\centering}\gnumbox{2009}
	&\gnumericPB{\centering}\gnumbox{7175}
	&\gnumericPB{\centering}\gnumbox{967999}
	&\gnumericPB{\centering}\gnumbox{3274946}
	&\gnumericPB{\centering}\gnumbox{58.29\%}
	&\gnumericPB{\centering}\gnumbox{41.71\%}
\\
	 \gnumericPB{\centering}\gnumbox{2010}
	&\gnumericPB{\centering}\gnumbox{7848}
	&\gnumericPB{\centering}\gnumbox{1033298}
	&\gnumericPB{\centering}\gnumbox{3545223}
	&\gnumericPB{\centering}\gnumbox{57.01\%}
	&\gnumericPB{\centering}\gnumbox{42.99\%}
\\
	 \gnumericPB{\centering}\gnumbox{2011}
	&\gnumericPB{\centering}\gnumbox{8095}
	&\gnumericPB{\centering}\gnumbox{1102155}
	&\gnumericPB{\centering}\gnumbox{3824334}
	&\gnumericPB{\centering}\gnumbox{55.45\%}
	&\gnumericPB{\centering}\gnumbox{44.55\%}
\\
	 \gnumericPB{\centering}\gnumbox{2012}
	&\gnumericPB{\centering}\gnumbox{8253}
	&\gnumericPB{\centering}\gnumbox{1158923}
	&\gnumericPB{\centering}\gnumbox{4095396}
	&\gnumericPB{\centering}\gnumbox{53.91\%}
	&\gnumericPB{\centering}\gnumbox{46.09\%}
\\
	 \gnumericPB{\centering}\gnumbox{2013}
	&\gnumericPB{\centering}\gnumbox{8335}
	&\gnumericPB{\centering}\gnumbox{1228952}
	&\gnumericPB{\centering}\gnumbox{4362849}
	&\gnumericPB{\centering}\gnumbox{52.31\%}
	&\gnumericPB{\centering}\gnumbox{47.69\%}
\\
	 \gnumericPB{\centering}\gnumbox{2014}
	&\gnumericPB{\centering}\gnumbox{8490}
	&\gnumericPB{\centering}\gnumbox{1270887}
	&\gnumericPB{\centering}\gnumbox{4584087}
	&\gnumericPB{\centering}\gnumbox{50.83\%}
	&\gnumericPB{\centering}\gnumbox{49.17\%}
\\
	 \gnumericPB{\centering}\gnumbox{2015}
	&\gnumericPB{\centering}\gnumbox{8660}
	&\gnumericPB{\centering}\gnumbox{1326996}
	&\gnumericPB{\centering}\gnumbox{4854865}
	&\gnumericPB{\centering}\gnumbox{49.34\%}
	&\gnumericPB{\centering}\gnumbox{50.66\%}
\\
	 \gnumericPB{\centering}\gnumbox{2016}
	&\gnumericPB{\centering}\gnumbox{8734}
	&\gnumericPB{\centering}\gnumbox{1370378}
	&\gnumericPB{\centering}\gnumbox{5117270}
	&\gnumericPB{\centering}\gnumbox{48.18\%}
	&\gnumericPB{\centering}\gnumbox{51.82\%}
\\
	 \gnumericPB{\centering}\gnumbox{2017}
	&\gnumericPB{\centering}\gnumbox{8860}
	&\gnumericPB{\centering}\gnumbox{1398003}
	&\gnumericPB{\centering}\gnumbox{5318208}
	&\gnumericPB{\centering}\gnumbox{47.19\%}
	&\gnumericPB{\centering}\gnumbox{52.81\%}
\\
\end{longtable}

\ifthenelse{\isundefined{\languageshorthands}}{}{\languageshorthands{\languagename}}
\gnumericTableEnd
}
\end{center}

\vspace{-1 cm}
\singlespacing
\noindent Table 2. {\bf Number of multi-authored papers published in the last decade.} Number of scientific articles signed by no more than $i=1, 2, ... 20$ authors and published in journals included in the Science Citation Index Expanded.
\begin{center}
  {\renewcommand\normalsize{\scriptsize}
    \normalsize
\def\ifundefined#1{\expandafter\ifx\csname#1\endcsname\relax}

\ifundefined{inputGnumericTable}

	\def\gnumericTableEnd{\end{document}}


	\documentclass[12pt%
                    ]{report}
       \usepackage[latin1]{inputenc}
       \usepackage{fullpage}
       \usepackage{color}
       \usepackage{array}
       \usepackage{longtable}
       \usepackage{calc}
       \usepackage{multirow}
       \usepackage{hhline}
       \usepackage{ifthen}

	\begin{document}

\else

    \def\gnumericTableEnd{}


\fi


\providecommand{\gnumericmathit}[1]{#1} 
\providecommand{\gnumericPB}[1]%
{\let\gnumericTemp=\\#1\let\\=\gnumericTemp\hspace{0pt}}
 \ifundefined{gnumericTableWidthDefined}
        \newlength{\gnumericTableWidth}
        \newlength{\gnumericTableWidthComplete}
        \newlength{\gnumericMultiRowLength}
        \global\def\gnumericTableWidthDefined{}
 \fi
 \ifthenelse{\isundefined{\languageshorthands}}{}{\languageshorthands{english}}
\providecommand\gnumbox{\makebox[0pt]}

\setlength{\bigstrutjot}{\jot}
\setlength{\extrarowheight}{\doublerulesep}

\setlongtables

\setlength\gnumericTableWidth{%
	117pt+%
	53pt+%
	53pt+%
	53pt+%
	53pt+%
	53pt+%
	53pt+%
	53pt+%
	53pt+%
	53pt+%
	53pt+%
0pt}
\def\gumericNumCols{11}
\setlength\gnumericTableWidthComplete{\gnumericTableWidth+%
         \tabcolsep*\gumericNumCols*2+\arrayrulewidth*\gumericNumCols}
\ifthenelse{\lengthtest{\gnumericTableWidthComplete > \linewidth}}%
         {\def\gnumericScale{\ratio{\linewidth-%
                        \tabcolsep*\gumericNumCols*2-%
                        \arrayrulewidth*\gumericNumCols}%
{\gnumericTableWidth}}}%
{\def\gnumericScale{1}}


\ifthenelse{\isundefined{\gnumericColA}}{\newlength{\gnumericColA}}{}\settowidth{\gnumericColA}{\begin{tabular}{@{}p{117pt*\gnumericScale}@{}}x\end{tabular}}
\ifthenelse{\isundefined{\gnumericColB}}{\newlength{\gnumericColB}}{}\settowidth{\gnumericColB}{\begin{tabular}{@{}p{53pt*\gnumericScale}@{}}x\end{tabular}}
\ifthenelse{\isundefined{\gnumericColC}}{\newlength{\gnumericColC}}{}\settowidth{\gnumericColC}{\begin{tabular}{@{}p{53pt*\gnumericScale}@{}}x\end{tabular}}
\ifthenelse{\isundefined{\gnumericColD}}{\newlength{\gnumericColD}}{}\settowidth{\gnumericColD}{\begin{tabular}{@{}p{53pt*\gnumericScale}@{}}x\end{tabular}}
\ifthenelse{\isundefined{\gnumericColE}}{\newlength{\gnumericColE}}{}\settowidth{\gnumericColE}{\begin{tabular}{@{}p{53pt*\gnumericScale}@{}}x\end{tabular}}
\ifthenelse{\isundefined{\gnumericColF}}{\newlength{\gnumericColF}}{}\settowidth{\gnumericColF}{\begin{tabular}{@{}p{53pt*\gnumericScale}@{}}x\end{tabular}}
\ifthenelse{\isundefined{\gnumericColG}}{\newlength{\gnumericColG}}{}\settowidth{\gnumericColG}{\begin{tabular}{@{}p{53pt*\gnumericScale}@{}}x\end{tabular}}
\ifthenelse{\isundefined{\gnumericColH}}{\newlength{\gnumericColH}}{}\settowidth{\gnumericColH}{\begin{tabular}{@{}p{53pt*\gnumericScale}@{}}x\end{tabular}}
\ifthenelse{\isundefined{\gnumericColI}}{\newlength{\gnumericColI}}{}\settowidth{\gnumericColI}{\begin{tabular}{@{}p{53pt*\gnumericScale}@{}}x\end{tabular}}
\ifthenelse{\isundefined{\gnumericColJ}}{\newlength{\gnumericColJ}}{}\settowidth{\gnumericColJ}{\begin{tabular}{@{}p{53pt*\gnumericScale}@{}}x\end{tabular}}
\ifthenelse{\isundefined{\gnumericColK}}{\newlength{\gnumericColK}}{}\settowidth{\gnumericColK}{\begin{tabular}{@{}p{53pt*\gnumericScale}@{}}x\end{tabular}}

\begin{longtable}[c]{%
	b{\gnumericColA}%
	b{\gnumericColB}%
	b{\gnumericColC}%
	b{\gnumericColD}%
	b{\gnumericColE}%
	b{\gnumericColF}%
	b{\gnumericColG}%
	b{\gnumericColH}%
	b{\gnumericColI}%
	b{\gnumericColJ}%
	b{\gnumericColK}%
	}

	 \gnumericPB{\centering}\gnumbox{\textbf{\# articles by:}}
	&\gnumericPB{\centering}\gnumbox{\textbf{2008}}
	&\gnumericPB{\centering}\gnumbox{\textbf{2009}}
	&\gnumericPB{\centering}\gnumbox{\textbf{2010}}
	&\gnumericPB{\centering}\gnumbox{\textbf{2011}}
	&\gnumericPB{\centering}\gnumbox{\textbf{2012}}
	&\gnumericPB{\centering}\gnumbox{\textbf{2013}}
	&\gnumericPB{\centering}\gnumbox{\textbf{2014}}
	&\gnumericPB{\centering}\gnumbox{\textbf{2015}}
	&\gnumericPB{\centering}\gnumbox{\textbf{2016}}
	&\gnumericPB{\centering}\gnumbox{\textbf{2017}}
\\
	 \gnumericPB{\centering}\gnumbox{\textbf{1 author}}
	&\gnumericPB{\centering}\gnumbox{73808}
	&\gnumericPB{\centering}\gnumbox{74699}
	&\gnumericPB{\centering}\gnumbox{75568}
	&\gnumericPB{\centering}\gnumbox{74636}
	&\gnumericPB{\centering}\gnumbox{72706}
	&\gnumericPB{\centering}\gnumbox{71087}
	&\gnumericPB{\centering}\gnumbox{68511}
	&\gnumericPB{\centering}\gnumbox{66815}
	&\gnumericPB{\centering}\gnumbox{65178}
	&\gnumericPB{\centering}\gnumbox{62498}
\\
	 \gnumericPB{\centering}\gnumbox{\textbf{$\le$ 2 authors}}
	&\gnumericPB{\centering}\gnumbox{224878}
	&\gnumericPB{\centering}\gnumbox{231238}
	&\gnumericPB{\centering}\gnumbox{237675}
	&\gnumericPB{\centering}\gnumbox{240647}
	&\gnumericPB{\centering}\gnumbox{240413}
	&\gnumericPB{\centering}\gnumbox{240937}
	&\gnumericPB{\centering}\gnumbox{237165}
	&\gnumericPB{\centering}\gnumbox{236058}
	&\gnumericPB{\centering}\gnumbox{232985}
	&\gnumericPB{\centering}\gnumbox{228850}
\\
	 \gnumericPB{\centering}\gnumbox{\textbf{$\le$ 3 authors}}
	&\gnumericPB{\centering}\gnumbox{391368}
	&\gnumericPB{\centering}\gnumbox{407505}
	&\gnumericPB{\centering}\gnumbox{422235}
	&\gnumericPB{\centering}\gnumbox{434066}
	&\gnumericPB{\centering}\gnumbox{439167}
	&\gnumericPB{\centering}\gnumbox{447095}
	&\gnumericPB{\centering}\gnumbox{444570}
	&\gnumericPB{\centering}\gnumbox{447216}
	&\gnumericPB{\centering}\gnumbox{447805}
	&\gnumericPB{\centering}\gnumbox{443948}
\\
	 \gnumericPB{\centering}\gnumbox{\textbf{$\le$ 4 authors}}
	&\gnumericPB{\centering}\gnumbox{538245}
	&\gnumericPB{\centering}\gnumbox{564238}
	&\gnumericPB{\centering}\gnumbox{589092}
	&\gnumericPB{\centering}\gnumbox{611093}
	&\gnumericPB{\centering}\gnumbox{624712}
	&\gnumericPB{\centering}\gnumbox{642833}
	&\gnumericPB{\centering}\gnumbox{646027}
	&\gnumericPB{\centering}\gnumbox{654702}
	&\gnumericPB{\centering}\gnumbox{660279}
	&\gnumericPB{\centering}\gnumbox{659724}
\\
	 \gnumericPB{\centering}\gnumbox{\textbf{$\le$ 5 authors}}
	&\gnumericPB{\centering}\gnumbox{652270}
	&\gnumericPB{\centering}\gnumbox{687901}
	&\gnumericPB{\centering}\gnumbox{722446}
	&\gnumericPB{\centering}\gnumbox{755205}
	&\gnumericPB{\centering}\gnumbox{776687}
	&\gnumericPB{\centering}\gnumbox{805935}
	&\gnumericPB{\centering}\gnumbox{815932}
	&\gnumericPB{\centering}\gnumbox{834279}
	&\gnumericPB{\centering}\gnumbox{845160}
	&\gnumericPB{\centering}\gnumbox{848867}
\\
	 \gnumericPB{\centering}\gnumbox{\textbf{$\le$ 6 authors}}
	&\gnumericPB{\centering}\gnumbox{736860}
	&\gnumericPB{\centering}\gnumbox{780314}
	&\gnumericPB{\centering}\gnumbox{822926}
	&\gnumericPB{\centering}\gnumbox{864534}
	&\gnumericPB{\centering}\gnumbox{895138}
	&\gnumericPB{\centering}\gnumbox{934300}
	&\gnumericPB{\centering}\gnumbox{951937}
	&\gnumericPB{\centering}\gnumbox{978605}
	&\gnumericPB{\centering}\gnumbox{995837}
	&\gnumericPB{\centering}\gnumbox{1003703}
\\
	 \gnumericPB{\centering}\gnumbox{\textbf{$\le$ 7 authors}}
	&\gnumericPB{\centering}\gnumbox{792343}
	&\gnumericPB{\centering}\gnumbox{841197}
	&\gnumericPB{\centering}\gnumbox{889773}
	&\gnumericPB{\centering}\gnumbox{939086}
	&\gnumericPB{\centering}\gnumbox{976128}
	&\gnumericPB{\centering}\gnumbox{1023502}
	&\gnumericPB{\centering}\gnumbox{1047355}
	&\gnumericPB{\centering}\gnumbox{1081325}
	&\gnumericPB{\centering}\gnumbox{1103269}
	&\gnumericPB{\centering}\gnumbox{1116058}
\\
	 \gnumericPB{\centering}\gnumbox{\textbf{$\le$ 8 authors}}
	&\gnumericPB{\centering}\gnumbox{829702}
	&\gnumericPB{\centering}\gnumbox{882154}
	&\gnumericPB{\centering}\gnumbox{935310}
	&\gnumericPB{\centering}\gnumbox{990260}
	&\gnumericPB{\centering}\gnumbox{1032466}
	&\gnumericPB{\centering}\gnumbox{1086613}
	&\gnumericPB{\centering}\gnumbox{1115059}
	&\gnumericPB{\centering}\gnumbox{1154773}
	&\gnumericPB{\centering}\gnumbox{1181896}
	&\gnumericPB{\centering}\gnumbox{1198092}
\\
	 \gnumericPB{\centering}\gnumbox{\textbf{$\le$ 9 authors}}
	&\gnumericPB{\centering}\gnumbox{853888}
	&\gnumericPB{\centering}\gnumbox{909020}
	&\gnumericPB{\centering}\gnumbox{965507}
	&\gnumericPB{\centering}\gnumbox{1024224}
	&\gnumericPB{\centering}\gnumbox{1070302}
	&\gnumericPB{\centering}\gnumbox{1128542}
	&\gnumericPB{\centering}\gnumbox{1161147}
	&\gnumericPB{\centering}\gnumbox{1204739}
	&\gnumericPB{\centering}\gnumbox{1236128}
	&\gnumericPB{\centering}\gnumbox{1254763}
\\
	 \gnumericPB{\centering}\gnumbox{\textbf{$\le$ 10 authors}}
	&\gnumericPB{\centering}\gnumbox{870687}
	&\gnumericPB{\centering}\gnumbox{927821}
	&\gnumericPB{\centering}\gnumbox{986747}
	&\gnumericPB{\centering}\gnumbox{1048033}
	&\gnumericPB{\centering}\gnumbox{1097294}
	&\gnumericPB{\centering}\gnumbox{1158671}
	&\gnumericPB{\centering}\gnumbox{1193574}
	&\gnumericPB{\centering}\gnumbox{1240850}
	&\gnumericPB{\centering}\gnumbox{1274854}
	&\gnumericPB{\centering}\gnumbox{1296180}
\\
	 \gnumericPB{\centering}\gnumbox{\textbf{$\le$ 11 authors}}
	&\gnumericPB{\centering}\gnumbox{881065}
	&\gnumericPB{\centering}\gnumbox{939696}
	&\gnumericPB{\centering}\gnumbox{1000216}
	&\gnumericPB{\centering}\gnumbox{1063458}
	&\gnumericPB{\centering}\gnumbox{1114245}
	&\gnumericPB{\centering}\gnumbox{1177799}
	&\gnumericPB{\centering}\gnumbox{1214514}
	&\gnumericPB{\centering}\gnumbox{1263871}
	&\gnumericPB{\centering}\gnumbox{1299822}
	&\gnumericPB{\centering}\gnumbox{1322448}
\\
	 \gnumericPB{\centering}\gnumbox{\textbf{$\le$ 12 authors}}
	&\gnumericPB{\centering}\gnumbox{888280}
	&\gnumericPB{\centering}\gnumbox{947840}
	&\gnumericPB{\centering}\gnumbox{1009344}
	&\gnumericPB{\centering}\gnumbox{1074136}
	&\gnumericPB{\centering}\gnumbox{1126327}
	&\gnumericPB{\centering}\gnumbox{1191349}
	&\gnumericPB{\centering}\gnumbox{1229326}
	&\gnumericPB{\centering}\gnumbox{1280264}
	&\gnumericPB{\centering}\gnumbox{1317667}
	&\gnumericPB{\centering}\gnumbox{1341144}
\\
	 \gnumericPB{\centering}\gnumbox{\textbf{$\le$ 13 authors}}
	&\gnumericPB{\centering}\gnumbox{892744}
	&\gnumericPB{\centering}\gnumbox{952939}
	&\gnumericPB{\centering}\gnumbox{1015247}
	&\gnumericPB{\centering}\gnumbox{1080910}
	&\gnumericPB{\centering}\gnumbox{1134327}
	&\gnumericPB{\centering}\gnumbox{1200507}
	&\gnumericPB{\centering}\gnumbox{1239372}
	&\gnumericPB{\centering}\gnumbox{1291214}
	&\gnumericPB{\centering}\gnumbox{1329877}
	&\gnumericPB{\centering}\gnumbox{1354124}
\\
	 \gnumericPB{\centering}\gnumbox{\textbf{$\le$ 14 authors}}
	&\gnumericPB{\centering}\gnumbox{895749}
	&\gnumericPB{\centering}\gnumbox{956465}
	&\gnumericPB{\centering}\gnumbox{1019339}
	&\gnumericPB{\centering}\gnumbox{1085814}
	&\gnumericPB{\centering}\gnumbox{1139990}
	&\gnumericPB{\centering}\gnumbox{1206906}
	&\gnumericPB{\centering}\gnumbox{1246378}
	&\gnumericPB{\centering}\gnumbox{1299073}
	&\gnumericPB{\centering}\gnumbox{1338727}
	&\gnumericPB{\centering}\gnumbox{1363672}
\\
	 \gnumericPB{\centering}\gnumbox{\textbf{$\le$ 15 authors}}
	&\gnumericPB{\centering}\gnumbox{897905}
	&\gnumericPB{\centering}\gnumbox{959041}
	&\gnumericPB{\centering}\gnumbox{1022337}
	&\gnumericPB{\centering}\gnumbox{1089235}
	&\gnumericPB{\centering}\gnumbox{1144095}
	&\gnumericPB{\centering}\gnumbox{1211520}
	&\gnumericPB{\centering}\gnumbox{1251474}
	&\gnumericPB{\centering}\gnumbox{1304948}
	&\gnumericPB{\centering}\gnumbox{1345063}
	&\gnumericPB{\centering}\gnumbox{1370543}
\\
	 \gnumericPB{\centering}\gnumbox{\textbf{$\le$ 16 authors}}
	&\gnumericPB{\centering}\gnumbox{899352}
	&\gnumericPB{\centering}\gnumbox{960852}
	&\gnumericPB{\centering}\gnumbox{1024407}
	&\gnumericPB{\centering}\gnumbox{1091749}
	&\gnumericPB{\centering}\gnumbox{1146946}
	&\gnumericPB{\centering}\gnumbox{1214830}
	&\gnumericPB{\centering}\gnumbox{1255187}
	&\gnumericPB{\centering}\gnumbox{1309123}
	&\gnumericPB{\centering}\gnumbox{1349826}
	&\gnumericPB{\centering}\gnumbox{1375728}
\\
	 \gnumericPB{\centering}\gnumbox{\textbf{$\le$ 17 authors}}
	&\gnumericPB{\centering}\gnumbox{900437}
	&\gnumericPB{\centering}\gnumbox{962104}
	&\gnumericPB{\centering}\gnumbox{1025935}
	&\gnumericPB{\centering}\gnumbox{1093523}
	&\gnumericPB{\centering}\gnumbox{1149041}
	&\gnumericPB{\centering}\gnumbox{1217286}
	&\gnumericPB{\centering}\gnumbox{1257977}
	&\gnumericPB{\centering}\gnumbox{1312168}
	&\gnumericPB{\centering}\gnumbox{1353387}
	&\gnumericPB{\centering}\gnumbox{1379444}
\\
	 \gnumericPB{\centering}\gnumbox{\textbf{$\le$ 18 authors}}
	&\gnumericPB{\centering}\gnumbox{901213}
	&\gnumericPB{\centering}\gnumbox{963094}
	&\gnumericPB{\centering}\gnumbox{1027123}
	&\gnumericPB{\centering}\gnumbox{1094968}
	&\gnumericPB{\centering}\gnumbox{1150550}
	&\gnumericPB{\centering}\gnumbox{1219191}
	&\gnumericPB{\centering}\gnumbox{1260005}
	&\gnumericPB{\centering}\gnumbox{1314618}
	&\gnumericPB{\centering}\gnumbox{1356007}
	&\gnumericPB{\centering}\gnumbox{1382345}
\\
	 \gnumericPB{\centering}\gnumbox{\textbf{$\le$ 19 authors}}
	&\gnumericPB{\centering}\gnumbox{901806}
	&\gnumericPB{\centering}\gnumbox{963762}
	&\gnumericPB{\centering}\gnumbox{1028034}
	&\gnumericPB{\centering}\gnumbox{1096027}
	&\gnumericPB{\centering}\gnumbox{1151782}
	&\gnumericPB{\centering}\gnumbox{1220603}
	&\gnumericPB{\centering}\gnumbox{1261640}
	&\gnumericPB{\centering}\gnumbox{1316455}
	&\gnumericPB{\centering}\gnumbox{1358224}
	&\gnumericPB{\centering}\gnumbox{1384633}
\\
	 \gnumericPB{\centering}\gnumbox{\textbf{$\le$ 20 authors}}
	&\gnumericPB{\centering}\gnumbox{902332}
	&\gnumericPB{\centering}\gnumbox{964418}
	&\gnumericPB{\centering}\gnumbox{1028847}
	&\gnumericPB{\centering}\gnumbox{1096902}
	&\gnumericPB{\centering}\gnumbox{1152800}
	&\gnumericPB{\centering}\gnumbox{1221776}
	&\gnumericPB{\centering}\gnumbox{1262971}
	&\gnumericPB{\centering}\gnumbox{1317998}
	&\gnumericPB{\centering}\gnumbox{1360019}
	&\gnumericPB{\centering}\gnumbox{1386621}
\\
	 \gnumericPB{\centering}\gnumbox{\textbf{any \# authors}}
	&\gnumericPB{\centering}\gnumbox{905349}
	&\gnumericPB{\centering}\gnumbox{967999}
	&\gnumericPB{\centering}\gnumbox{1033298}
	&\gnumericPB{\centering}\gnumbox{1102155}
	&\gnumericPB{\centering}\gnumbox{1158923}
	&\gnumericPB{\centering}\gnumbox{1228952}
	&\gnumericPB{\centering}\gnumbox{1270887}
	&\gnumericPB{\centering}\gnumbox{1326996}
	&\gnumericPB{\centering}\gnumbox{1370378}
	&\gnumericPB{\centering}\gnumbox{1398003}
\\
\end{longtable}

\ifthenelse{\isundefined{\languageshorthands}}{}{\languageshorthands{\languagename}}
\gnumericTableEnd
}
\end{center}

\doublespacing
\section*{Methods.}
\noindent Data were obtained from inCites (https://incites.thomsonreuters.com/\#/analytics; ``Journals'' and ``People''), an integrated web-based platform based on the data from the Web of Science by Clarivate Analytics (provided by the University of Zurich). We focused on the journals included in the Science Citation Index Expanded which covers 177 subject areas out of the 252 subject categories comprised by the Web of Science schema (i.e., ``Filters: Research Area''; we selected the areas from http://mjl.clarivate.com/scope/scope\_scie/). We count as papers only regular scientific articles that report original research, not books, reviews, editorials, letters to the editor and the like (i.e., ``Filters: Document type''; we selected ``Article''). We restrict our search to the journals listed in the Journal Citation Reports (i.e., ``Thresholds: JIF Quatile''; we selected Q1, Q2, Q3, Q4). The number of papers published by $i=1, 2, ... n$ authors was obtained using an additional threshold (i.e., ``Thresholds: Authors per Document''; we selected Min=1 and Max=1 for single-authored papers, ...).

\end{document}